\documentclass[sigconf]{acmart}

\usepackage[english]{babel}
\usepackage{blindtext}
\usepackage{algorithm}
\usepackage{algorithmicx}
\usepackage{algpseudocode}
\usepackage{multirow}
\usepackage{wasysym}
\usepackage{mathtools}
\usepackage{amsthm} 
\usepackage{marvosym}
\usepackage{enumitem}

\usepackage{booktabs} 
\usepackage{makecell}
\usepackage[compact]{titlesec}
\titlespacing\section{0pt}{6pt plus 4pt minus 2pt}{0pt plus 2pt minus 2pt}
\titlespacing\subsection{0pt}{6pt plus 4pt minus 2pt}{0pt plus 2pt minus 2pt}
\titlespacing\subsubsection{0pt}{6pt plus 4pt minus 2pt}{0pt plus 2pt minus 2pt}
\usepackage[font=small,format=plain,labelfont=bf,textfont=it]{caption}

\renewcommand\footnotetextcopyrightpermission[1]{} 

\renewcommand\footnotetextcopyrightpermission[1]{} 
\setcopyright{none}

\acmDOI{}

\acmISBN{}

\acmConference[arXiv]{}
\acmYear{22 Feb 2023}

\acmPrice{}
\begin{document}
	\title[AC-DC]{AC-DC: Adaptive Ensemble Classification \\ for Network Traffic Identification}
	
	
	
	\author{Xi Jiang}
	\affiliation{%
		\institution{University of Chicago}
	}
	\author{Shinan Liu}
	\affiliation{%
	\institution{University of Chicago}
	}
	\author{Saloua Naama}
	\affiliation{%
	\institution{Université Savoie Mont Blanc}
	}
	\author{Francesco Bronzino}
	\affiliation{%
	\institution{École Normale Supérieure de Lyon}
	}
	\author{Paul Schmitt}
	\affiliation{%
	\institution{University of Hawaii, Manoa}
	}
	\author{Nick Feamster}
	\affiliation{%
	\institution{University of Chicago}
	}
	
\renewcommand{\paragraph}[1]{\vspace*{0.03in}\noindent\textbf{#1}}
\newcommand{\shinan}[1]{\textcolor{blue}{[#1 - Shinan]}}
\renewcommand{\shortauthors}{Jiang.et al.}
\newcommand{\eg}{{\it e.g.}}
\newcommand{\ie}{{\it i.e.}}
\newcommand{\etal}{{\it et al.}}
\newcommand{\sysname}{{AC-DC}}

\begin{sloppypar}

\begin{abstract} 
	Accurate and efficient network traffic classification is important for many
	network management tasks, from traffic prioritization to anomaly detection.
	Although classifiers using pre-computed flow statistics (\eg, packet
	sizes, inter-arrival times) can be efficient, they may experience lower accuracy
	than techniques based on raw traffic, including packet captures.
	Past work on representation learning-based classifiers applied to network traffic
	captures has shown to be more accurate, but slower and requiring considerable
	additional memory resources, due to the substantial costs in feature preprocessing.
	In this paper, we explore this trade-off and develop the Adaptive Constraint-Driven
	Classification (\sysname) framework to efficiently curate a pool of classifiers
	with different target requirements, aiming to provide comparable classification
	performance to complex packet-capture classifiers while adapting to varying network 
	traffic load.

	\sysname{} uses an adaptive scheduler that tracks current
	system memory availability and incoming traffic rates to determine the optimal
	classifier and batch size to maximize classification
	performance given memory and processing constraints. Our evaluation shows that
	{\sysname} improves classification performance by more than 100\% compared to
	classifiers that rely on flow statistics alone; compared to
	the state-of-the-art packet-capture classifiers,
	{\sysname} achieves comparable performance (less than 12.3\% lower
	in F1-Score), but processes traffic over 150x faster.

\end{abstract}

\maketitle

\section{Introduction}\label{sec:intro}

Network traffic classification is a common practice in network management that
involves identifying the services and applications being used on the Internet
infrastructure. Traffic classification enables network operators to carry out
many crucial network operations: ISPs leverage traffic classification to
monitor bandwidth utilization to prioritize certain subsets of traffic (\eg, for
latency-sensitive applications) to ensure QoS or avoid network congestion, for capacity and 
resource planning, or to detect malicious traffic, and
more~\cite{singhal2013state, jiang2007lightweight,
5504793,roesch2005snort,sommer2003bro,paxson1999bro,stewart2005architecture,baker2004cisco}.
Efficient, accurate traffic classification is thus critical for effective
network operations.

Network traffic classification research has a long history dating back decades.
Early traffic classification solutions relied on heuristics developed by domain
experts who derived the necessary features to infer traffic
categories~\cite{4351725, sen2004accurate,moore2005toward}. Beyond their classification 
performance, these methods were extremely efficient as a result of
the simple features and traffic signatures they used. More explicitly, the total
amount of time they required to preprocess extracted traffic data, feed
the processed data to the classification method, and provide a classification answer
was extremely low. We refer to this amount of time as the \emph{time-to-decision
(TTD)}, \ie, the end-to-end time from incoming traffic to a classifier decision. 
Unfortunately, as the Internet evolves (and, in particular, as traffic is increasingly encrypted) and 
consolidates (\ie, services are operated by fewer infrastructure providers), many of these classifiers have become less accurate
and difficult to maintain as the engineered features become unavailable or less
informative and quickly out of date.

\begin{figure}[t]
	\centering     
	\includegraphics[width=\columnwidth]{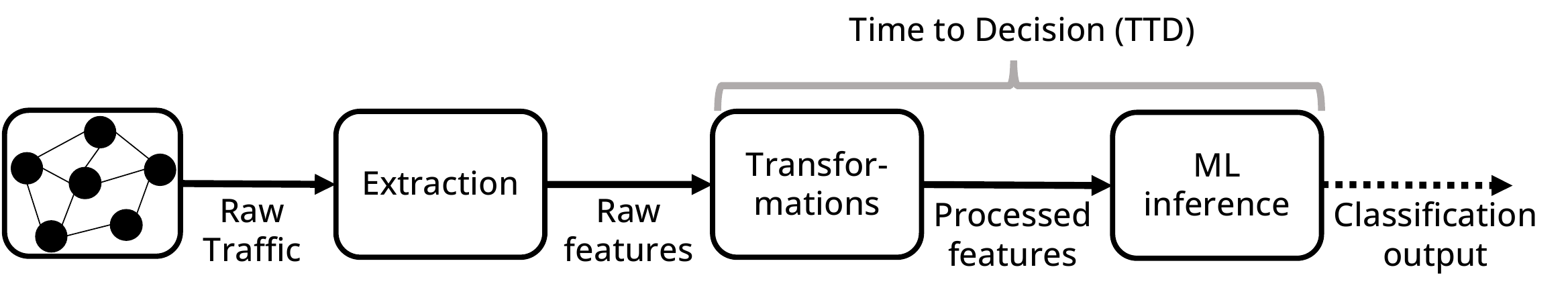}
	\caption{Typical phases of a network traffic classification 
	pipeline and the range of processing time that TTD includes.}
	\label{fig:ttddef}
\end{figure}

Recent advances in machine learning techniques have enabled new
classification methods that overcome some of the limitations of heuristic
solutions. A first class of ML-based classification models are
flow-statistics-based
methods~\cite{paxson1994empirically,dewes2003analysis,claffy1995internet,
lang2003synthetic,lang2004synthetic,bernaille2006early,karagiannis2005blinc}.
Similar to heuristic-based methods, these classifiers often rely strictly on
engineered features and have the advantage of being fast at feature computation,
making them good candidates to handle high network traffic volume with fast
classification speed and minimal memory overhead. Unfortunately, they also share
the limitations of previous methods, suffering from deteriorated performance
upon traffic pattern changes and the Internet architecture evolutions at large.

To improve the classification performance, \ie, model accuracy, of existing
solutions and overcome the problems of features availability due to the
increasing prevalence of encrypted traffic payloads,
new solutions that employ representation learning-based methods have been
developed to classify traffic based directly from raw packet
captures\cite{rimmer2017automated,zheng2022mtt,
lotfollahi2020deep,akbari2021look,yao2019identification,wang2020encrypted,
shapira2019flowpic,cui2019session,bu2020encrypted,ma2021encrypted,sun2020encrypted}.
A common characteristic (and often, drawback) of these classifiers is that
they pay less attention to
their practicality in handling realistic network traffic volumes under resource
constraints. Thus, despite their high classification performance,
they cannot
satisfy the scale and bandwidth requirements of modern networks, becoming
irrelevant against faster traffic rates. 

To date, we are aware of no existing approach that can achieving high classification
performance metrics coupled with the ability to meet the resource demands imposed
by the larger networks they must be deployed within. In this paper, we address this
critical trade-off, by considering the practical deployment setting of a traffic
classifier (as shown in Figure~\ref{fig:ttddef}). We recognize that a classifier
consists of at least (1)~a preprocessing component that transforms raw features
into ML-usable input formats and (2)~an attached trained model that makes
inferences using the preprocessed data. First, we identify that the
efficiency bottlenecks of packet-capture classifiers often lie in the
preprocessing component, where large amounts of required raw traffic features
are in need of transformation, rather than the actual model inference stage.
Existing
approaches~\cite{koksal2022markov,crankshaw2017clipper,li2020train,qiu2022traffic,tong2014high}
often aim to overcome the trade-offs by enhancing packet-capture classifiers'
throughput via tactics such as adjusting the complexity of the learning
algorithm and the feature representation used. Although these strategies may
improve the time-complexity of the model execution component of the TTD, the
preprocessing of selected features remains the main bottleneck. Furthermore, the
high memory overhead associated with implementing these classifiers is often
neglected, which is also a critical consideration for practical deployment.

In this paper, we focus on tackling the feature preprocessing bottleneck and
propose the \textbf{Adaptive Constraint-Driven Classification (\sysname)}
framework. \sysname{} takes an ensemble approach to provide a balance between
classification performance and efficiency. \sysname{} achieves high classifier performance while meeting system requirements by implementing three key contributions: 
\begin{enumerate}[leftmargin=*,itemsep=0pt,topsep = 0pt]
	\item Rather than yielding a single classifier that demands fixed features,
	\sysname{} maintains a pool of classifiers with different feature requirements
	to allow for variations in the classification speed and memory utilization.
	This approach diverges from conventional ensemble practices in machine
	learning, where multiple algorithms are utilized to address potential
	variations in data distributions. Instead, the focus is on providing flexible
	choices for the feature requirements to adapt to traffic volumes and accommodate varying system
	resource constraints.
	\item \sysname{} includes a new heuristic-based feature exploration algorithm
	that not only optimizes for high classification performance, but also
	prioritizes low system overhead (\eg, memory efficiency), producing feature combinations that
	exhibit the best trade-off between performance and efficiency. Our algorithm
	uses approximation techniques to quickly identify the most relevant features,
	enabling the generation of a more robust and efficient set of classifiers.
	\item Finally, \sysname{} computes and preserves the classification
	performance, TTD, and memory utilization for each classifier at various
	\emph{batch sizes} (\ie, the number of flows to collect and extract features
	from) before ingesting the flows and executing an instance of the classifier.
	Facing different traffic rates and memory availability on a given link,
	\sysname{} implements an adaptive scheduler that leverages the measured
	information to select the optimal classifier and batch size that (1) satisfies
	the current memory availability and traffic rate, and (2) provides the ideal
	balance between performance and efficiency.
\end{enumerate}

We evaluate \sysname{} on a dataset of traffic flows spanning a
variety of applications and services. Our results indicate that \sysname{} 
remains robust under various traffic rates and memory
constraints. Meanwhile, it consistently balances the trade-off between
model performance and efficiency by outperforming conventional packet-capture
classifiers by orders of magnitude in terms of throughput while largely
preserving classification performance that is substantially higher than classifiers based on flow statistics.

\section{Motivation}\label{sec:motivation}
In this section, we illustrate the inherent trade-offs that there exist across
performance and efficiency for network traffic classification techniques. We
first introduce the classification task we focus on, together with existing
solutions: flow-statistics and packet-capture classifiers, selecting for
each class a representative solution from the state-of-the-art. We then identify
the efficiency bottlenecks of these classifiers and show that they
largely occur in raw data preprocessing step rather than in the model
execution/inference stages.

\subsection{Traffic Classification Background}

\paragraph{The Classification Task.}
Network traffic classification is a continuous process that
commences with the interception of Internet traffic and ends with
grouping the traffic into predefined categories,
\eg, normal/abnormal traffic or specific services/applications.
This paper focuses on the task of application identification,
which involves classifying network traffic at the flow level
into corresponding applications, such as YouTube and Netflix.
Figure~\ref{fig:ttddef} illustrates the crucial steps
involved in a classifier. These steps encompass initial operations,
such as extracting raw features from the captured traffic, and more
machine learning (ML)-oriented procedures, including converting these
raw features into a format suitable for ML and using trained models
to perform inference on flows with the processed features. Our focus
is on appraising the performance and efficiency of the ML components
of the classifier.


\paragraph{Flow-statistics-based.}
Flow-statistics classifiers are ML-based approaches for traffic
classification, which are often
inspired by previous work developed using heuristics-based techniques. These
classifiers typically involve the use of computed statistics from traffic flows
to make predictions, \eg, throughput, packet inter-arrival times, etc. 
We focus on and evaluate a representative flow-statistics classifier -
Gaussian Mixture Models clustering (GMM) - proposed by Bernaille
\etal~\cite{bernaille2006early} to perform the classification task using the
packet sizes and inter-arrival times of the first four non-zero size packets of
the TCP connections of traffic flows.

\paragraph{Packet-capture-based.}
Recent ML-based traffic classifiers often make use of increased computation power,
ingesting raw traffic contents (\ie, packet captures) as inputs. 
This approach allows a more in-depth analysis of network traffic, thereby
providing a comprehensive understanding of hidden patterns and characteristics
that can aid in accurate traffic classification. To this end, we select for the
same classification task using a packet-capture-based on a one-dimensional
convolutional neural network (CNN) classifier, as described by Lotfollahi
\etal~\cite{lotfollahi2020deep}. In this classifier, only the first three
packets of each flow are provided as inputs. This occurs because 
learning from the initial three packets from the flows already allows for the
classifier to
achieve F1-Scores well above 0.9. We follow the original design and feed the
first 1500 bytes (including the IP header) of each packet into the classifier
and, as the CNN initially performs packet-level classification, we
take the majority vote of the three packet-level predictions to arrive at a
flow-level classification decision.

\subsection{Performance and Efficiency Trade-off}
We conduct an assessment of the selected classifiers by training and testing
them on a curated dataset consisted of near 20,000 individual flows spanning
across ten prominent applications. The dataset is divided into train/test sets
using a 50/50 split, with a relatively larger test set compared to the
conventional 80/20 split. We perform the split in this way because our evaluation emphasizes
not only performance but also the efficiency of the classifiers during the
classification process, which demands a sufficiently large test set.

We vary the traffic rate from 10 to 7,000 flows/second by randomly sampling
specified numbers of flows from the test set and examine classifiers'
performance and efficiency in terms of weighted F1-Score, TTD, and in-use
memory. \textit{Traffic rate} in this study indicates the number of flows per
second that the system is able to capture and extract raw features for. As these
conventional classifiers do not elaborate on the explicit number of flows that
are required to be captured before initiating a new instance of the classifier,
\ie, the batch size, we follow the assumption that a new instance of the
classifier is initiated every second to classify all traffic received in the
previous second.


\begin{table}[thb]
	\resizebox{\columnwidth}{!}{
		\begin{tabular}{|l|l|l|l|}
			\hline
			\multirow{2}{*}{\textbf{\makecell{Classifier\\Type}} } &  \multicolumn{1}{l|}{\textbf{\makecell{Performance}}} & \multicolumn{2}{l|}{\textbf{\makecell{Efficiency}}}\\
			\cline{2-4}
			&  \textbf{\makecell{F1-Score}}& \textbf{\makecell{TTD\\(seconds)}} & \textbf{\makecell{Memory\\Requirement\\(GB)}} \\ \hline
			
			Flow-statistics  & 0.391 & 0.023&0.503\\ 
			&  & (76.3\% feature preprocessing)&\\
			&  & (23.7\% model execution)&\\\hline
			Packet-capture  & 0.977  & 19.689&82.463\\
			&  & (74.7\% feature preprocessing)&\\
			&  &    (25.3\% model execution)&\\\hline
	\end{tabular}}
	\caption{The flow-statistic classifier excels in efficiency but the packet-capture classifier shows significantly
		better classification performance.}
	
	\label{tab:overalltradeoff}
\end{table}
\paragraph{\textbf{Model Performance.}} As illustrated in
Table~\ref{tab:overalltradeoff}, and consistent with previous work,
our evaluation shows that the flow-statistics GMM classifier has an average F1-Score of 0.391, significantly lower than the packet-capture-based classifier with an average F1-Score of 0.977.
This result is expected because
the model cannot fit the complex underlying distribution of the data using the few provided features.
The features used by the flow-statistics classifier also become
less informative, which also reduces the classification performance.
For example, the size information is becoming less distinctive:
for the second packet, 56.27\% of flows across 10 applications share similar sizes around 1500 bytes
(detailed comparison in Figure~\ref{fig:gmmdistribution} in Appendix).
On the contrary, packet-capture
classifiers
tend to exhibit higher classification performance,
due to the abundance of learnable features and the capability to automatically
identify informative features without the need for manual feature selection.




\paragraph{Time to Decision.} 
Table~\ref{tab:overalltradeoff} shows that the packet-capture classifier
exhibits average TTD that is $\sim$855x higher than the flow-statistics-based
GMM. To understand where the majority of time is spent, we split the TTD into
the proportion of time required to process the features vs the time to execute
the inference model. We observe that, for both classifiers, the TTD bottlenecks
are largely caused by high feature preprocessing time, rather than at the model
execution/inference steps. For both classifiers, the average model execution
time never exceeds $30\%$ of the total TTD. Although we report the
average results in these efficiency trade-off analysis, we also provide
fine-grained comparisons at all evaluated traffic rates which can be found in
Appendix (Figures~\ref{fig:ttd_compare_app} and \ref{fig:mem_compare_app}).
The observed trends across various
traffic rates remain relatively consistent as the reported average results.


%

\paragraph{In-Usage Memory.}
The memory requirement in this example refers to the additional memory consumption incurred
by the classifiers beyond the raw feature extraction, including storing
intermediate features during preprocessing, loading trained models for
classification, and generating final predictions. When dealing with a high
volume of traffic, it is expected to have multiple instances of the classifiers
running to mitigate delays and potential buffer overflow. In this paper, we
define the memory requirement of the classifier as the minimum system memory
necessary to support the concurrent execution of multiple instances of the
classifiers needed at the specified traffic rate (with details in
Section~\ref{sec:method}). As depicted in Table~\ref{tab:overalltradeoff}, on
average, the packet-capture-based CNN necessitates $\sim$163x more system memory
than the flow-statistics-based Gaussian Mixture Model. This difference is
anticipated as the feature space and computation complexity of the
packet-capture classifiers are significantly higher, resulting in substantial
increases in memory overhead. These results echo that generic packet-capture
classifiers might incur in memory violations if deployed on systems that are not
equipped with the required on-board memory.


Overall, considering the comparison results presented, we observe that there
exist trade-offs between flow-statistics and packet-capture classifiers in
terms of performance and efficiency. Our goal is to balance these trade-offs by
introducing a new traffic classification framework to retain high
performance of packet-capture classifiers while achieving better efficiency,
approaching that
of flow-statistics classifiers.

\section{Adaptive Constraint-Driven Classification}\label{sec:method}

\begin{figure*}[tb]
	\centering     
	\includegraphics[width=\textwidth]{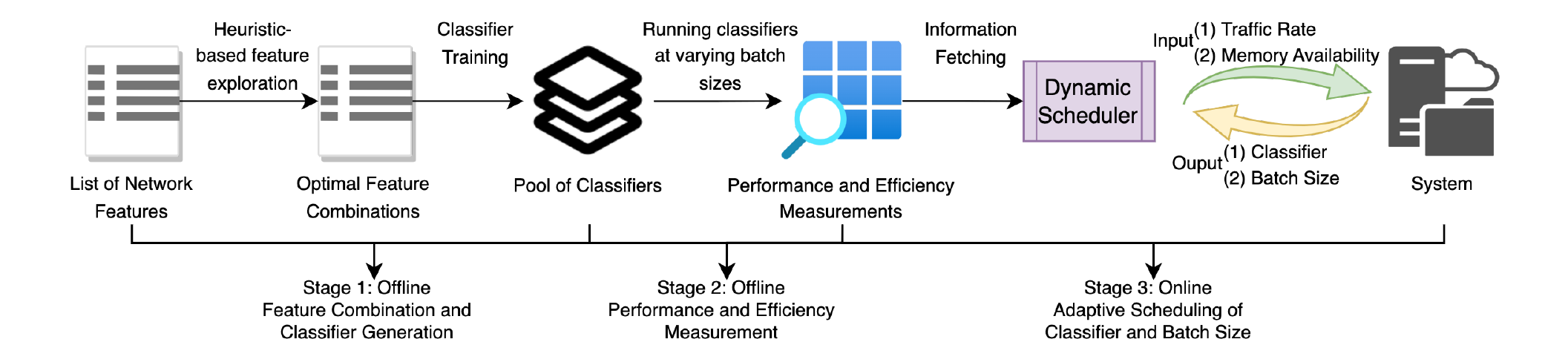}
	\caption{Adaptive Constraint-Driven Classification ({\sysname}) framework overview}
	\label{fig:module}
\end{figure*}

Our evaluation of existing classifiers shows that a significant portion of the
efficiency overhead in network traffic classifiers stems from the need to
preprocess raw features to attain high classification performance.
Nevertheless, when dealing with large volumes of network traffic, it may be
acceptable to make small sacrifices in performance to achieve higher
efficiency and ensure that the classifier can function properly. And, under
varying system constraints, it is crucial to balance performance and
efficiency without unnecessarily compromising on either metric.

To tackle this challenge, we propose the Adaptive Constraint-Driven Classification
(\sysname) framework. {\sysname} leverages an ensemble approach by curating a
pool of classifiers, each with varying feature requirements and, therefore,
different performance-efficiency trade-offs. The proposed method is in
accordance with the standard ML practices that use ensemble models for
network traffic classification. However, our ensemble approach 
is unique because it concentrates on modifying the feature specifications to adapt to
fluctuating system-resource limitations, instead of relying on a collection of ML
algorithms to accommodate diverse data distributions. This empowers us to make
an informed decision on selecting the most suitable classifier with different
feature requirements to implement, based on the computational resources
and traffic volume that are available at the time.

As illustrated in Figure~\ref{fig:module}, the framework comprises a
three-stage pipeline with the following components:
\begin{enumerate}[leftmargin=*,itemsep=0pt]
    \item An offline bootstrapping stage that uses a heuristic
    feature exploration algorithm to approximate the optimal feature
    combinations for a balance between performance and efficiency; this stage
    generates a pool of classifiers, each with different feature requirements as
    determined by the outcome of the feature exploration algorithm
    (Section~\ref{subsec:heuristic});
    \item An offline stage that measures the actual performance and
    efficiency of the classifiers at different batch sizes when deployed on the
    implementing system (Section~\ref{subsec:batches});
    \item An online classification stage which employs an adaptive
    scheduler to continually monitor current traffic rates and memory
    availability to select the most suitable classifier and batch size for
    classification (Section~\ref{subsec:scheduler}).
\end{enumerate}

In the rest of the section, we first discuss the performance and efficiency
metrics that are at the core of \sysname{}'s adaptive mechanisms, followed by
the details of each component.

\subsection{Performance and Efficiency Metrics}\label{subsec:metrics}
Before describing the details of the individual components of the framework,
we first establish clear definitions for the performance and efficiency metrics
used in the framework.

\paragraph{Classification Performance.} In this paper, we compute all
classifiers' performance using the F1-Score, which is a harmonic mean of
precision and recall. It is worth noting that this performance metric can be
replaced with any other desired metric, depending on the specific application
targets, \eg, prioritize recall to avoid false negatives avoidance.

\paragraph{Time-to-Decision.}
As shown in Figure~\ref{fig:ttddef}, we define TTD as the total elapsed time
from the initiation of the feature preprocessing step to the conclusion of the
model inference step, that is, when classification decisions have been made on
all flows. TTD is an important metric for classification frameworks when deployed in
networking systems: large TTDs can force systems to lose packets or not have the
required information in time. In this work, to obtain accurate TTD measurements,
we remove all unnecessary outputs and functionalities in the classifiers, such
as progress logging and metadata storage. To reflect realistic time consumption,
I/O times for intermediate step, such as loading extracted raw features, are
included. In the case of {\sysname}, the classifier switching time is also
considered in the TTD. We do not consider raw feature extraction times 
in the TTD  because the feature-extraction process tends
to be relatively fast and the techniques used for extraction vary among network
system implementors, making it difficult to conduct a fair comparison across
different tools.

\paragraph{Memory Requirement.}
We do not measure memory requirements of the classifiers by
referring to their total memory allocations under no memory constraint because
such allocations reflect the \textit{ideal} but not 
\textit{minimum}
memory requirements. Instead, we define the memory requirement as the minimum
amount of RAM and SWAP memory that is necessary for the classifier instance to
function normally. This includes ensuring that (1)~the classification processes
are not terminated prematurely and (2) the TTD does not increase due to
insufficient memory availability. When multiple instances of the
classifier are running in parallel, we calculate the aggregated minimum memory
requirement to support concurrent execution of all instances. To determine such
minimum memory requirements, we use Linux memory control
groups (\texttt{cgroup}) to impose limitations on the amount of memory available to the
classifier instance. Subsequently, we use a binary search algorithm to
iteratively reduce memory usage to the smallest amount that does not cause
the instance to behave abnormally.

\subsection{Heuristics-based Feature Exploration and Classifier Generation}\label{subsec:heuristic}

%

The performance and efficiency of network classifiers stems as much from the
model in use as from the features that are fed to it. In the first stage of
its pipeline, \sysname{} selects candidate features across a large list of
available ones to generate a pool of candidate classification models.

\paragraph{Preliminary Feature Selection.}
To carry out any feature exploration, we first need to establish an
initial set of network-level features. We also need to find a tool for
selective preprocessing of specified subsets of the features from raw traffic
captures into machine-learning-compatible representations. In this specific
implementation of {\sysname}, we define the features using the header fields of
the flow packets, \eg, IPV4\_TTL, TCP\_OPT, etc, and use
nPrint~\cite{holland2021new} to generate the required network data
representations. This is because nPrint supports an easily accessible packet
header field subscription interface. In this representation, each packet of the
flows is encoded into a normalized, binary format preserving the underlying
semantics of each packet. For future implementations, one can easily substitute
nPrint with any desired network traffic encoder without disrupting the
functionality of the \sysname{} framework.

We define 37 individual features, which include the IPV4, TCP, and UDP
header fields as well as the packet payload. And we proceed to remove the packet
payload because all testing traffic are TLS encrypted and the payload should
only contain noise that does not contribute to the classification task. We also
remove all noticeable fields that may contribute to model over-fitting,
including the IP addresses and ports for both ends of the connections. Only the
header fields from the first three packets in each flow are used as they already
produce sufficiently good performance. After this preliminary feature selection,
we arrive at an initial list of 32 features that are available to use for
feature exploration and generating the pool of classifiers. Detailed features
are available in Appendix (Table~\ref{tab:feature_selection}).

\paragraph{Heuristic-Based Feature Exploration.}
Once candidate features are selected, the framework identifies feature combinations
to create candidate classifiers with high (1) performance-to-TTD and (2)
performance-to-memory requirement ratios. The goal is to identify candidate
classifiers that strike a good balance between performance and
efficiency. Traditional feature selection techniques are not ideal for this task
as they are primarily optimized for identifying small feature subsets with high
performance, while ignoring efficiency considerations. Additionally, they
typically eliminate correlated features. However, in our scenario, correlated
features may still be appropriate if they improve classification performance
with minimal efficiency overhead.

A naive approach is to explore all possible feature combinations and select
the top feature subsets that yield the highest ratios. However, this would
entail a massive feature exploration space that requires us to train and
evaluate $\sum_{k=1}^{33} \binom{33}{k}$ classifiers, which is impractical.
While there exists some non-exhaustive mechanisms for this purpose, such as
hill
climbing~\cite{goswami2019filter,skalak1994prototype,abualigah2017feature,farmer2004large},
these approaches still require training/testing/evaluating a large number of intermediate
classifiers during the process of computing the optimal feature subsets. And
they are sometimes faraway from ideal solutions due to challenges such as
oscillations in local optima.

To address this challenge, we rely on two observations: (1)~When training the
model that includes all 33 features and looking at the permutation shuffling
feature importance, we observe a strong correlation (coefficient of 0.753)
between feature importance and the marginal improvement in performance achieved
by making each feature available to the classifier. (2)~We observe that there
exists a strong correlation (coefficient of above 0.99) between the aggregated
number of bits in a feature subset and the corresponding efficiency overhead in
terms of TTD and memory requirements.

Given these observations, we can approximate the expected performance and
efficiency changes when incorporating a feature into the model's feature set.
Thus, we can abstract the previous mentioned optimization ratios by considering
the ratio between the aggregated feature importance and number of bits for any
given feature subsets. We leverage these findings and construct a
heuristic-based algorithm to generate optimal feature subsets that works as
follows:
\begin{itemize}[leftmargin=*,itemsep=0pt]
    \item For each feature, we compute its permutation feature importance
    ($FI_i$) and total number of bits $Bits_i$. We then calculate the ratio
    $\frac{FI_i}{Bits_i}$ and rank features based on the calculated ratio. We show a
    list of example features after this step in
    Appendix (Table~\ref{tab:final_feature_selection}).
    \item We then determine the size of the desired pool using two parameters:
    $sizes$, a set of feature requirement sizes to consider; and $num_{combos}$,
    number of different combinations to generate at each feature requirement
    size. Thus, the total number of classifiers is $pool\_size = len(sizes)
    \cdot num_{combos}$.
    \item Finally, for each feature requirement size (n) in the set of $sizes$,
    we iteratively find $num_{combos}$ distinct combinations of $n$ features
    with the highest sum of $\frac{FI_i}{Bits_i}$ ratios. This procedure yields
    a set of feature subsets with size equal to $pool\_size$ and they are
    subsequently used as distinct feature requirements to generate the pool of
    classifiers.
\end{itemize}

The heuristic-based algorithm we have developed offers several advantages over the exhaustive
search approach: Firstly, its time complexity is bounded by the desired
$pool\_size$, which results in a significantly reduced computational load
compared to exhaustive search. Secondly, the proposed heuristic algorithm
eliminates the need for training/testing/evaluating excessive amounts of
classifiers during the feature subset derivation process, thereby further
reducing the computational burden. Finally, the computed feature subsets are
also optimized toward the performance-to-efficiency ratio per the strong
correlations discovered through the observations.

\paragraph{Classifier Generation.}
We then use the computed feature subsets as the feature requirements to generate
the classifiers. The goal of this paper is to improve the performance and
efficiency of traffic classification by manipulating the feature space and batch
size. Hence, the specific choice of the ML algorithm is less significant, as
long as it is consistent with the baseline. In this paper, we choose LightGBM as
an example model to provide comparable results to the nPrintml-based LightGBM
baseline introduced later in Section~\ref{sec:evaluation}. For each feature
requirement, we use nPrint to preprocess the specified features and then feed
them into the training of a LightGBM model to complete the generation of the
classifier.

\subsection{Performance and Efficiency Measurements at Varying Batch Sizes}\label{subsec:batches}
Upon obtaining the pool of classifiers created at the previous stage, we gather
detailed metrics on their performance and efficiency by executing them at
various batch sizes. We enforce the concept of batch sizes by storing extracted
raw features from the test set for a number of flows equivalent to the batch
size into a designated location. Then, classifier instances are initiated to
preprocess and make predictions on all the raw features in a single batch. 

\paragraph{Why Vary the Batch Size?}
In general machine learning, the batch size is a hyperparameter that determines
the number of samples to gather before initializing an instance of the
classifier to do a bulk preprocessing and classification on all input samples.
In our context, the batch size refers to the number of traffic flows to collect
raw features  from before calling the classifier instance. As shown in
Section~\ref{sec:motivation} where we simply set the batch size equal to the
traffic rate, increasing batch sizes can greatly impact the TTD and memory
requirements, particularly for packet-capture classifiers. If the concept of
batch size is not taken into account and classifier are ran periodically, for
example, once per second, the associated efficiency overhead can become
excessively high when the traffic rate is high. As a result, we collect
measurements on the classifiers at varying batch sizes during the offline
bootstrapping phase which can subsequently be utilized by the adaptive scheduler
to make informed decisions when selecting the most appropriate classifier and
determining the optimal batch size. Prior
research~\cite{crankshaw2017clipper,lim2019packet,zheng2020learning} have
attempted to optimize batch size for reducing model execution time. However,
they do not place a significant emphasis on the adaptive adjustment of batch
sizes to optimize the overall TTD and do not account for the impact that varying
batch sizes may have on memory overhead.

\paragraph{Memory Requirement on Concurrent Instances of Classifiers.}
With a pre-determined batch size, it is possible to have multiple instances of a
classifier running in parallel. This occurs when the TTD of a classifier
instance is greater than the time required to fill up the batch size but enough
system memory is available to execute concurrent instances of the model. To
determine the memory requirement necessary to execute two or more models
concurrently, we evaluate required memory as the aggregate memory needed to
support concurrent execution of all the classifier instances without any failure
or delay.
%

Provided with any batch size $B$ and classifier $C$, we can measure the (1)
end-to-end time $(TTD_C(B))$ it takes for the classifier to arrive at the
prediction and (2) the unit memory requirement $(m_C(B))$ for running one
instance of $C$. Sequentially, given batch size $B_t$ and traffic rate
$R_t$ at time $T_t$, we can formulate the maximum possible number of concurrently running
instances of any classifier $C$ with $TTD_C$ as
\begin{equation}
	\label{eqn:1}
	N_t(B_t,R_t,TTD_C) = \lceil \frac{TTD_C(B_t)}{B_t/R_t} \rceil
\end{equation}
and the total memory requirement can be expressed as
\begin{equation}
	\label{eqn:2}
M_t(N_t,m_C) = N_t\cdot m_C
\end{equation}

Given any classifier instance with a trained model and a predetermined batch
size, we can calculate its total memory requirement according to the
above formulation. For example, in our evaluation of the existing classifiers
where the batch size is by default equal to the traffic rate at all time, their
total memory requirement can be simply expressed as $\lceil TTD_C(B_t=R_t)
\rceil \cdot m_C(B_t)$.

\subsection{Adaptive Scheduler}\label{subsec:scheduler}
Given the generated pool of classifiers and established their corresponding
measurements, we now introduce an adaptive scheduler that continuously observes
the incoming network traffic rate and memory availability, and outputs the
optimal classifier and batch size with the best balance between performance and
efficiency.

\paragraph{Optimization Goal.} 
Facing any traffic rate, the aim of the adaptive scheduler is to determine the
optimal combination of classifier and batch size that balances high performance
with minimal TTD. To translate this goal into a executable optimization
function, the scheduler employs an iterative approach to search through the
performance and efficiency measurements and return the combination of classifier
and batch size with the highest performance-to-TTD ratio.

\paragraph{Minimum Performance Requirement.} 
Because the scheduler is designed to locate the classifier and batch size that
yield to the highest performance-to-TTD ratio, it seeks for a balance
between the two metrics but does not necessarily guarantee a lower bound on the
final performance.
In practical deployments, implementing parties may
have a minimum acceptable performance that they require in their classification
tasks. As a result, we add the functionality to specify an \textit{minimum
performance requirement (MPR)} to the adaptive scheduler which proceeds to
filter out any combination that does not meet the requirement. When no combination
can meet the requirement, the scheduler defers to the one with the highest 
performance.

\begin{algorithm}[htb]
	\caption{Adaptive Scheduler}\label{alg:cap}
	\begin{algorithmic}
		\State \textbf{Input:} traffic rate, mem availability
		\State \textbf{Data:} performance/efficiency measurements
		\State $candidates \gets type(list)$
		\For{\texttt{combination in Data}}
		\If{total mem requirement for combination <= mem availability}
		\State candidates.append(combination)
		\EndIf
		\EndFor
		\If{len(candidates)!=0}
		\State optimal combination $\gets candidates[0]$
		\For{\texttt{candidate in candidates}}
		\If{candidate performance-to-TTD ratio < optimal combination performance-to-TTD ratio}
		\State optimal combination  = candidate
		\EndIf
		\EndFor
		\EndIf
		\State \textbf{Return} optimal combination
	\end{algorithmic}
\end{algorithm}

Algorithm~\ref{alg:cap} describes the functionality of the adaptive
scheduler after the MPR has been enforced.
Given any constraint on the system memory resource, the scheduler observes the
memory requirement of the combinations at the given traffic rate and filter out
ones that result in a violation of memory availability. For example, if the
memory availability is 5 GB and the traffic rate is 1500 flows/second, a combination
with a classifier TTD of 1.5 seconds, a batch size of 500 flows, and a unit
memory requirement of 1.5 GB is going to be eliminated as the total memory
requirement (per eq. (\ref{eqn:1}) and (\ref{eqn:2})) is $\lceil
\frac{1.5}{500/1500} \rceil\cdot 1.5 = 6.75$ GB which exceeds the availability.
After filtering out all such combinations, the scheduler conducts an iterative
search procedure to identify the combination with the highest performance-to-TTD
ratio. The resulting classifier and batch size determine how the
incoming traffic flows are classified. As the traffic rate and memory availability change,
the scheduler adjusts the chosen combination to optimize the described ratio.

\section{Evaluation}\label{sec:evaluation}

We evaluate the performance of \sysname{} against state-of-the-art classifiers.
All classifiers are instructed to perform flow-level application classification
on a carefully selected network traffic flow dataset. Our experiments are
designed to confirm if, under both memory-rich and memory-scarce environments,
{\sysname} balances performance and efficiency, showing
better classification results than the flow-statistics classifier while
maintaining classification throughput higher than conventional packet-capture
classifiers. We further explore the robustness of the framework by profiling its
behavior under varying traffic rates and memory availability.

\subsection{Evaluation Setup}
\begin{table}[thb]
	\centering
	\resizebox*{\columnwidth}{!}{
		\begin{tabular}{|l|l|l|l|}
			\toprule
			\textbf{Macro Services} & \textbf{Total Number}   & \textbf{Application Labels}& \textbf{Collection Date}  \\
			& \textbf{of Flows}    & \textbf{(Number of Flows)}&   \\
			\hline
			Video Streaming~\cite{bronzino2019inferring}                      & 9465             & Netflix (4104)     &   2018-06-01    \\
			&              & YouTube (2702)     &       \\
			&              & Amazon (1509)     &       \\
			&              & Twitch (1150)     &       \\ \hline
			Video Conferencing~\cite{macmillan2021measuring}                        & 6511             & MS Teams (3886)         &   2020-05-05    \\
			&              & Google Meet (1313)     &       \\
			&              & Zoom (1312)     &       \\ \hline
			Social Media                       & 3610              & Facebook (1477)           &   2022-02-08	   \\ 
			&              & Twitter (1260)     &       \\
			&              & Instagram (873)     &       \\
			\bottomrule
		\end{tabular}
	}
	\caption{Summary of the network traffic flows dataset. We evaluate a 10-class 
		classifier covering all applications across these three macro service types.}
	\label{tab:stats}
\end{table}
\paragraph{Dataset.}
There are many sources for network traces, such as the ISCX VPN-NonVPN traffic
dataset~\cite{draper2016characterization}, the CAIDA Anonymized Internet
Traces~\cite{caida_2019}, and the QUIC dataset~\cite{tong2018novel}. However,
these datasets are relatively old and may not accurately reflect the current
Internet topology, particularly concerning the studied services. Additionally,
many of these datasets do not have clearly defined labels for the associated
applications or services at the flow level, which is essential for evaluating
classifier performance. As a result, we decide to evaluate all classifiers the
curated dataset described in Table~\ref{tab:stats} which consists of
\textit{video streaming}~\cite{bronzino2019inferring}, \textit{video
conferencing}~\cite{macmillan2021measuring}, and \textit{social media} traffic
flows spanning ten different major applications and a wide range of collection
dates. The traffic is captured as pcap files and cleaned by (1)~examining the
DNS queries to identify IP addresses relevant to the services/applications,
(2)~filtering out irrelevant traffic to contain only traffic associated with the
applications/services using the identified IP addresses, and (3)~separating
traffic into individual flows using the 5-tuple attributes (source and
destination IP, source and destination port, and the protocol field). We use the
application labels, such as Netflix or YouTube, for each preprocessed flow to
evaluate classification performance.
These traffic from the above-mentioned datasets are all TLS encrypted
and we use the aggregation of them as the general dataset for evaluating
{\sysname} and the baseline classifiers\footnote{All traces used in this paper
are sanitized and contain no personally identifiable information (PII).}.

\paragraph{Testbed Setup.}
All experiments are conducted on a Ubuntu 20.04.4 LTS operating
system with a Linux 5.4.0-135-generic kernel and an x86-64 architecture capable
of running 64-bit software. The CPU is an AMD EPYC 7502P 32-Core Processor, with
64 threads, 32 cores per socket, and 2 threads per core. The system has
1 socket, 1 NUMA node, and runs at a base clock speed of 1.49GHz and boost
clock speed of 2.5GHz. The system offers high
flexibility to run both lightweight tasks and large, demanding workloads that
require many cores and threads, such as network traffic processing
and classification. All classifiers described in this study by default spread
their work loads across available cores for parallel computing.

\paragraph{Classification Throughput.}
\sysname{} optimizes the balance between performance and
efficiency. While performance can be represented using the F1-Score, it can be
difficult to grasp the efficiency-related time-to-decision. To better present
our results, we convert the TTD into \textit{classification throughput} for all
evaluated classifiers and batch sizes at different traffic rates and memory
availability. Classification throughput here is defined as the number of traffic
flows that a classifier can preprocess and make inference decisions on within
one second after raw feature extraction. A higher classification throughput
serves as an indicator of better time efficiency. For example, if raw features
are extracted for 7,000 flows in one second and the classifier is able to process
and make decisions on 5,000 of those flows in the subsequent second, then the
classification throughput would be 5,000 flows/second, 71.4\% of the
traffic rate.   
We can derive {\sysname}'s classification throughput based on the measured TTD
because the latter reflects the amount of time needed for the classification
framework to complete preprocessing and inferencing on all flows collected in
the previous second, \ie, lower TTD results in higher classification throughput.
Formally, the classification throughput can be calculated as
$\frac{traffic\:rate}{TTD}$.

\subsection{Performance Under No Memory Constraint}
We first examine the performance and efficiency of {\sysname} in resource-rich
environments with abundant memory availability. Our objective is to classify
network flows at varying traffic rates, which range from 100 to 15,000
flows/second. We emulate different traffic rates aggregating randomly sampled
flows from the test set. We utilize the \textit{flow-statistics classifier
(GMM)} to establish an upper bound for classification throughput because it is
capable of preprocessing and classifying flows at the same rate as the incoming
traffic within the range under evaluation. On the other hand, we use two
classifiers as baselines for classification throughput to represent
packet-capture classifiers. These include the previously examined
\textit{Deep-Packet CNN classifier}~\cite{lotfollahi2020deep} and a
\textit{nPrintml-based LightGBM classifier} that uses all 33 features. We
include the LightGBM classifier as an additional example baseline as it uses the
same preprocessing tool (nPrint) and ML model as the {\sysname} implementation
under evaluation. This enables us to perform a fair comparison with minimum bias
from model and preprocessing tool selection. We also include the CNN classifier
to demonstrate our classification framework can provide better performance and
efficiency trade-off compared to packet-capture classifiers with different
preprocessing tools and ML models.

We consider feature requirement sizes ranging from one to nine with 10
different combinations at each size, resulting in a pool of 90 distinctive
classifiers each with different feature requirements. 
We set the maximum feature requirement size to nine, as increasing the maximum feature
requirement size beyond nine creates classifiers and feature sets that are
highly unlikely to be selected by the scheduler, \ie, features ranked 10th or
below (by the heuristic algorithm) start to deviate towards significantly high
overhead in face of low performance contribution. This is largely due to the
chosen feature granularity and, with finer-grained features, we can increase the
maximum feature requirement size accordingly. Example classifiers in this pool
are in the Appendix (Table~\ref{tab:evaluation_classifiers}).

\begin{figure}[t]
	\centering     
	\includegraphics[width=\columnwidth]{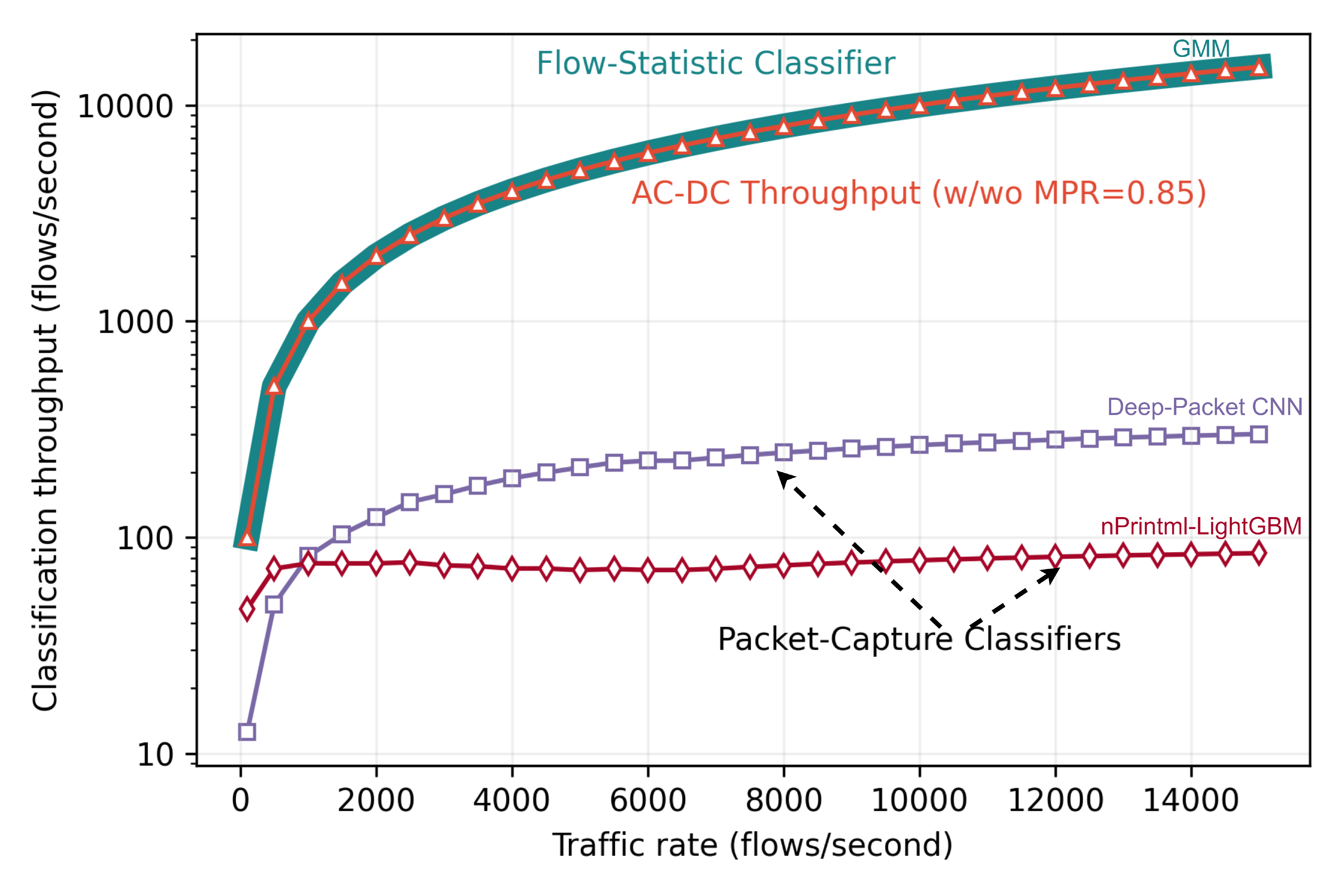}
	\caption{With no system memory constraint:
		{\sysname} consistently produces classification throughput as high as the upper bound
		and outperforms the conventional packet-capture classifiers.}
	\label{fig:eval_handled}
\end{figure}

\paragraph{\sysname{} Outperforms Packet-Capture Classifiers in Classification Throughput.}
We first study the achieved classification throughput across the different
classifiers. The results, shown in Figure~\ref{fig:eval_handled}, indicate that
the {\sysname} is able to match the throughput upper bound generated by the
flow-statistics-based GMM and outperform the conventional packet-capture-based
LightGBM and CNN by 176.95 and 48.98 times, as the traffic rate grows to 15,000
flows/second, respectively. We evaluate \sysname's classification throughput
both with and without the minimum performance requirement set at 0.85 in
F1-Score, and the framework meets the upper bound in both cases. In
this scenario, the framework delivers consistently high classification
throughput by selecting the classifier with the highest performance-to-TTD ratio
at a batch size of one and initiating as many instances of the chosen classifier
as required since no memory constraint is enforced. Thus, theoretically, the
{\sysname} consistently sustains a high classification
throughput given enough system memory, regardless of the increasing traffic
rate. 

The classification throughput, defined as the
number of flows that a classifier can preprocess and generate final predictions
for within a one-second time window following the extraction of raw features,
may not always reflect the actual TTD. Despite achieving 100\% traffic rate for
classification throughput, the actual TTD may be less than one second. As
observed from Table~\ref{tab:f1eval}, our framework achieves the upper bound
throughput but with a lower average TTD as compared to the flow-statistics
classifier. However, it still outperforms the packet-capture classifiers with a
substantially lower average TTD.

\begin{table}[t]
	\resizebox{0.85\columnwidth}{!}{
		\begin{tabular}{|l|l|l|l|l|l|l|}
			\hline
			\multirow{2}{*}{\textbf{\makecell{Classifier\\Type}} } & \multirow{2}{*}{\textbf{\makecell{ML\\Model}}}& \multirow{2}{*}{\textbf{\makecell{Processed \\ Features}}} & \multicolumn{3}{l|}{\textbf{\makecell{F1-Score}}} & \multirow{2}{*}{\textbf{\makecell{Avg. TTD\\(seconds)}}}\\
			\cline{4-6}
			& & &\textbf{\makecell{Avg.}}& \textbf{\makecell{Max.}} & \textbf{\makecell{Min.}} &\\ \hline
			
			Flow-statistics & GMM & Integers & 0.394 & 0.465&0.338&0.039\\ \hline
			\textbf{{\sysname}} & \textbf{LightGBM} & \textbf{standardized bits}  & \textbf{0.855}&\textbf{0.869}&\textbf{0.853}&\textbf{0.235}\\ \hline
			\multirow{2}{*}{Packet-capture} & LightGBM & standardized bits & 0.944 & 0.948&0.939&96.097\\
			\cline{2-7}
			& CNN & Sparse matrix & 0.975  & 0.988&0.961&30.486\\ \hline
	\end{tabular}}
	\caption{With no system memory constraint: {\sysname} outperforms
		flow-statistics classifier with 2x higher F1-Score and
		lower TTD compared to packet-capture classifiers.}
	\label{tab:f1eval}
\end{table}

\paragraph{{\sysname} Achieves Higher Classification Performance Than the
Flow-Statistics Classifier.} 
We now evaluate \sysname's classification performance against existing
classifiers. Overall, packet-capture classifiers have higher classification
performance, while the flow-statistics-based GMM get the lowest F1 score. We
first set the minimum performance requirement for {\sysname} at a F1-Score of
0.85 and the evaluation results. Table~\ref{tab:f1eval}, shows that {\sysname}
achieves an average F1-Score of 0.855, which is only 0.089 and 0.12 lower than
the upper bound performance of LightGBM and CNN, respectively. More importantly,
it outperforms the flow-statistics classifier by 117\% in F1-Score. When no
minimum performance requirement is specified, {\sysname} still reaches an
F1-Score of 0.8 which is 103\% higher than the flow-statistics classifier. 

These results highlights that, under no memory constraint, {\sysname} is capable
of outperforming packet-capture classifiers in terms of classification
throughput and flow-statistic classifiers in terms of classification
performance.

\subsection{Suitability for Memory-Constrained Environment}
Our previous results demonstrate that, with no memory constraint, {\sysname} 
achieves 100\% classification throughput with high performance for the
evaluated traffic rates. We proceed by confirming that the framework
can effectively maintain this balance even with low memory availability. We first
examine the minimum memory that {\sysname} requires for functioning and
\textit{maintaining a high 100\% throughput} that matches the
incoming traffic rate. In the case of the baseline packet-capture classifiers,
they experience low throughput even with unlimited memory availability as shown
in Figure~\ref{fig:eval_handled}. Thus, we measure their minimum memory
requirement to support the maximum throughput they are able to achieve. The
evaluation starts with a traffic rate of 500 flows/second which gradually
increasing up to 7,000 flows/second with a granularity of 500 flows/second.

\begin{figure}[t]
	\centering     
	\includegraphics[width=0.85\columnwidth]{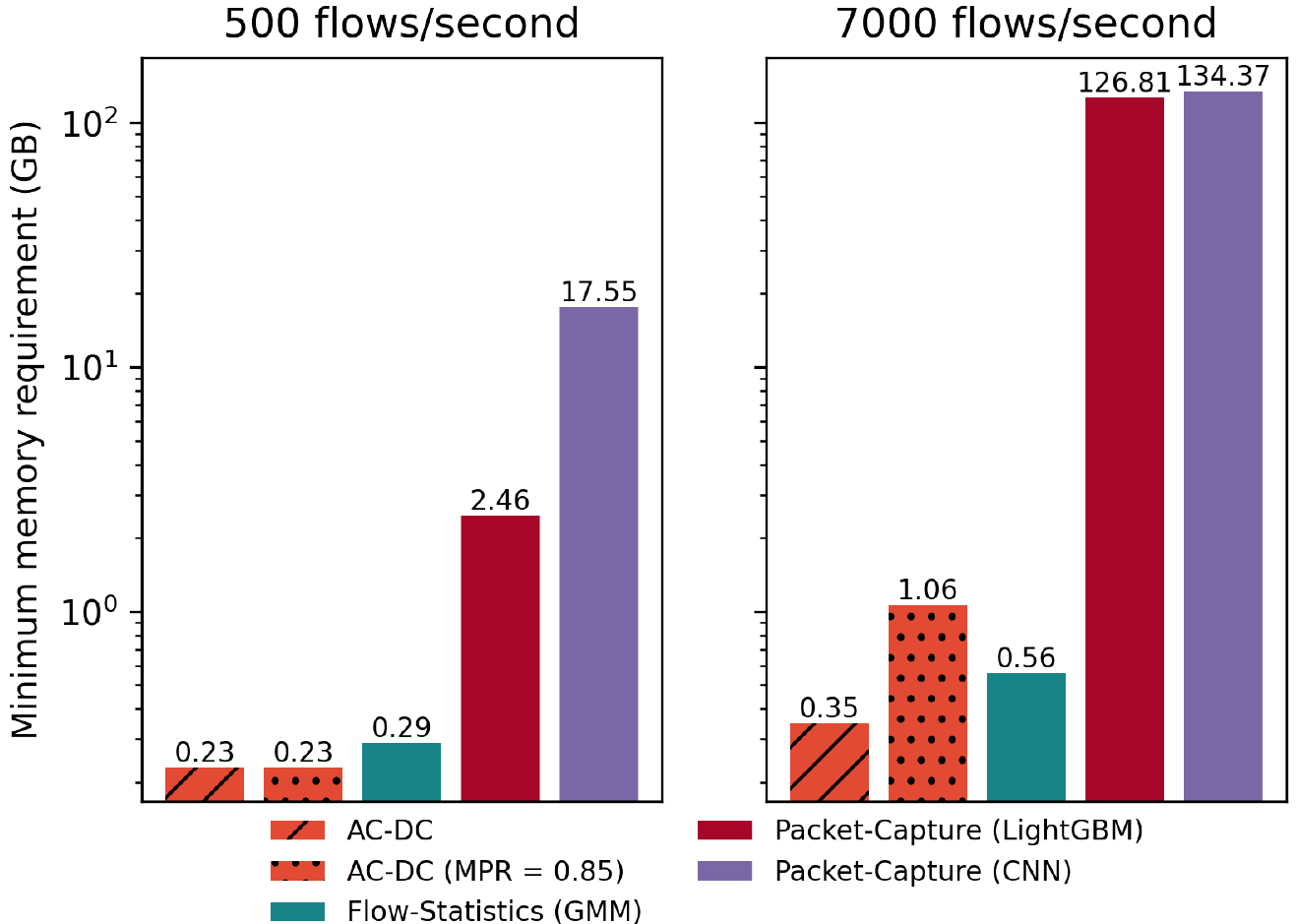}
	\caption{{\sysname} shows significantly lower minimum memory requirements than packet-capture
		classifiers for both low and high traffic rates.}
	\label{fig:eval_mem}
\end{figure}

\paragraph{{\sysname} Requires Less Memory Than Flow-Statistics Classifiers.} 
As demonstrated in Figure~\ref{fig:eval_mem}, the minimum memory requirements of
{\sysname} are lower than those of both flow-statistics and packet-capture
classifiers at the higher tested traffic rates (500 and 7,000 flows/second). In
absence of any performance constraints on the {\sysname} framework, it is
evident that {\sysname} is less prone to memory exhaustion and subsequent
failures. This characteristic of {\sysname} makes it a more suitable option for
deployments where memory availability is a limiting factor. This is primarily
due to heuristic-based classifier generation process, which aims to minimize TTD
and memory requirements for all generated classifiers. Consequently, regardless
of the classifier selected by the scheduler from the pool, the resulting memory
requirement will be relatively low. Fine-grained minimum memory requirements
can be found in Appendix (Figure~\ref{fig:eval_mem_compare_app}).

\paragraph{Imposing a Minimum Performance Requirement Increases Memory Requirements.}
Although the absence of minimum performance requirement on {\sysname} results in
minimized memory demands, it also leads to relatively low classification
performance, with an average F1-Score of 0.802. Thus, we further explore the
efficiency of the framework by imposing a minimum performance requirement of
0.85 in F1-Score and the result is also shown in Figure~\ref{fig:eval_mem}. With
the added constraint, we observe a slight increase in the overall minimum memory
requirement for maintaining 100\% throughput, which grows higher than the
flow-statistics classifier (starting at a traffic rate of 3,500 flows/second).
This behavior is expected, as higher classification performance necessitates the
selection of classifiers with more feature requirements, which occupy more
memory during preprocessing and classification. However, the resulting memory
requirements are still close to those of the flow-statistics classifier and
significantly lower than the packet-capture classifiers. At a traffic rate of
7,000 flows/second, {\sysname}'s minimum memory requirement is 118.8x and 126x
less than the nPrintml-LightGBM and Deep-Packet CNN classifiers, respectively.

\paragraph{{\sysname} Achieves Performance and Efficiency Balance, with 100\% Throughput at Minimal Memory.}
After demonstrating that the framework is better-suited for environments with
limited memory due to its relatively low memory requirements to sustain high
performance and throughput, we aim to establish that {\sysname} achieves
a better balance between performance and efficiency when compared to the
baseline classifiers. We analyze this by measuring the F1-Score and TTD of the
framework at varying traffic rates up to 15,000 flows/second. The framework is
only allowed to access the minimum amounts of memory needed to sustain a 100\%
classification throughput.

\begin{figure}[t]
	\centering     
	\includegraphics[width=\columnwidth]{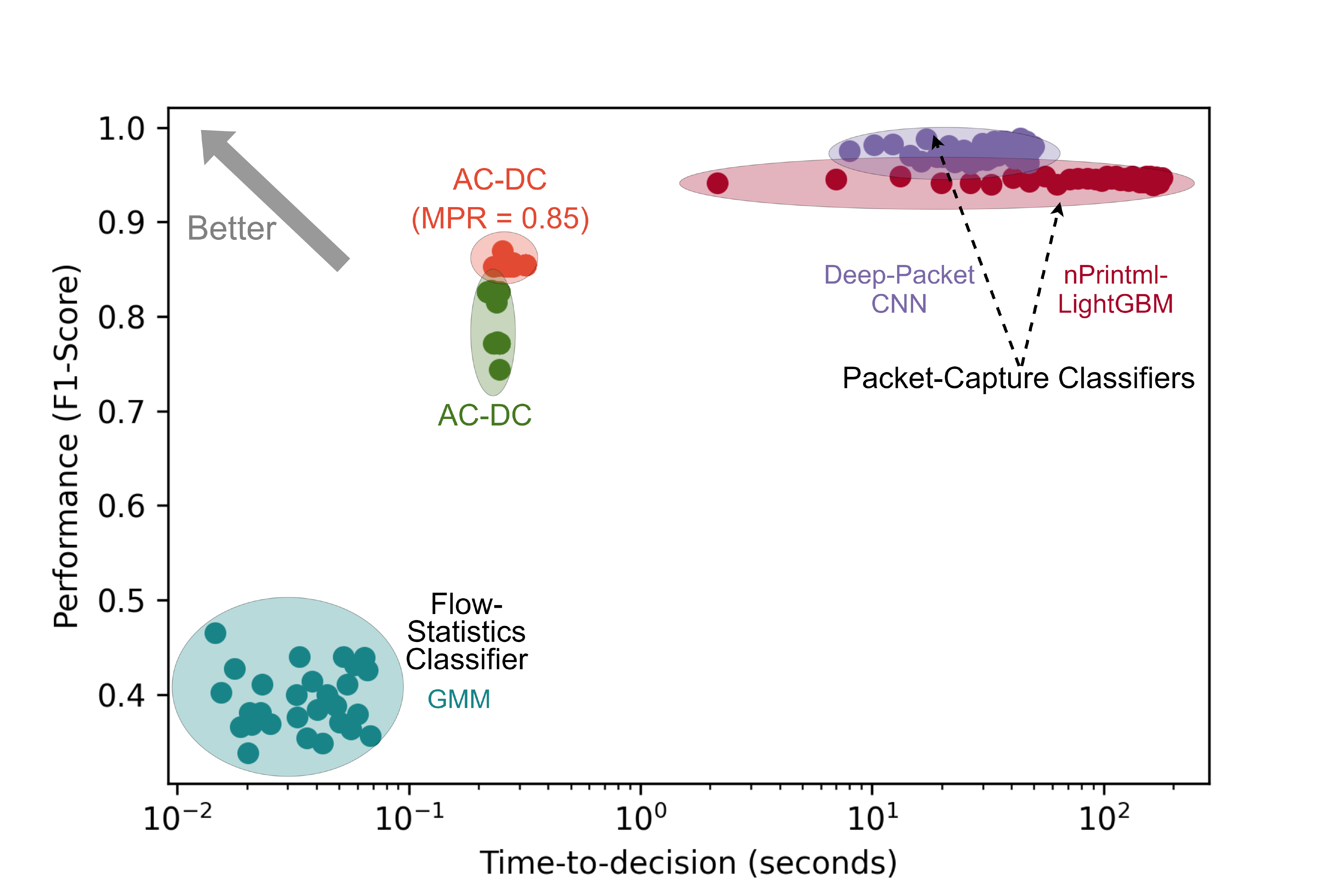}
	\caption{Classifier performance and efficiency at 100\% throughput and minimum memory requirements for varying traffic rates:
		{\sysname} shows more balanced trade-offs compared to both flow-statistics 
		and packet-capture classifiers.}
	\label{fig:tradeoffatmmr}
\end{figure}

The results, shown in Figure~\ref{fig:tradeoffatmmr}, reveal that \sysname{}
consistently balances between high F1-Score and low TTD (\ie, high
classification throughput). Unlike the baseline classifiers that typically
exhibit either high F1-Scores or low TTDs, but not both, {\sysname} is capable
of choosing a better combination of classifiers and batch sizes that yields both
high F1-Score and low TTD without severely compromising on either metric. This
is because the optimization goal enforced on our adaptive scheduler is designed
to produce outcomes that lead to the highest performance-to-TTD ratio, as
opposed to placing excessive emphasis on one metric over the other. At the same
time, our curated pool of classifiers and the corresponding feature requirements
are also pre-selected to maintain this balance.

\subsection{Behaviors Under Varying Memory Constraints and Traffic Rates}
Finally, we verify the behaviors under
changing memory constraints and traffic rates. This is to reflect realistic
deployment scenarios where traffic rates (and hence memory demands) change
during the course of a day.

\paragraph{{\sysname} Adjustments Facing Varying Memory Availability.}
To assess the influence of memory accessibility on \sysname's behavior, we
conduct experiments where we fix the traffic rate to 15,000 flows/second and
incrementally increase the memory available to {\sysname}. Note that the
selection of the fixed traffic rate is arbitrary as the goal of this
experimental setup is for us to isolate and discern the relationship between
memory availability and the framework's behavior. In all subsequent results, the
classification throughput maintains a consistent 100\% of the incoming traffic
rate. Experiments are conducted both with and without implementing a minimum
performance requirement. 

\begin{figure}[t]
	\centering     
	\includegraphics[width=0.85\columnwidth]{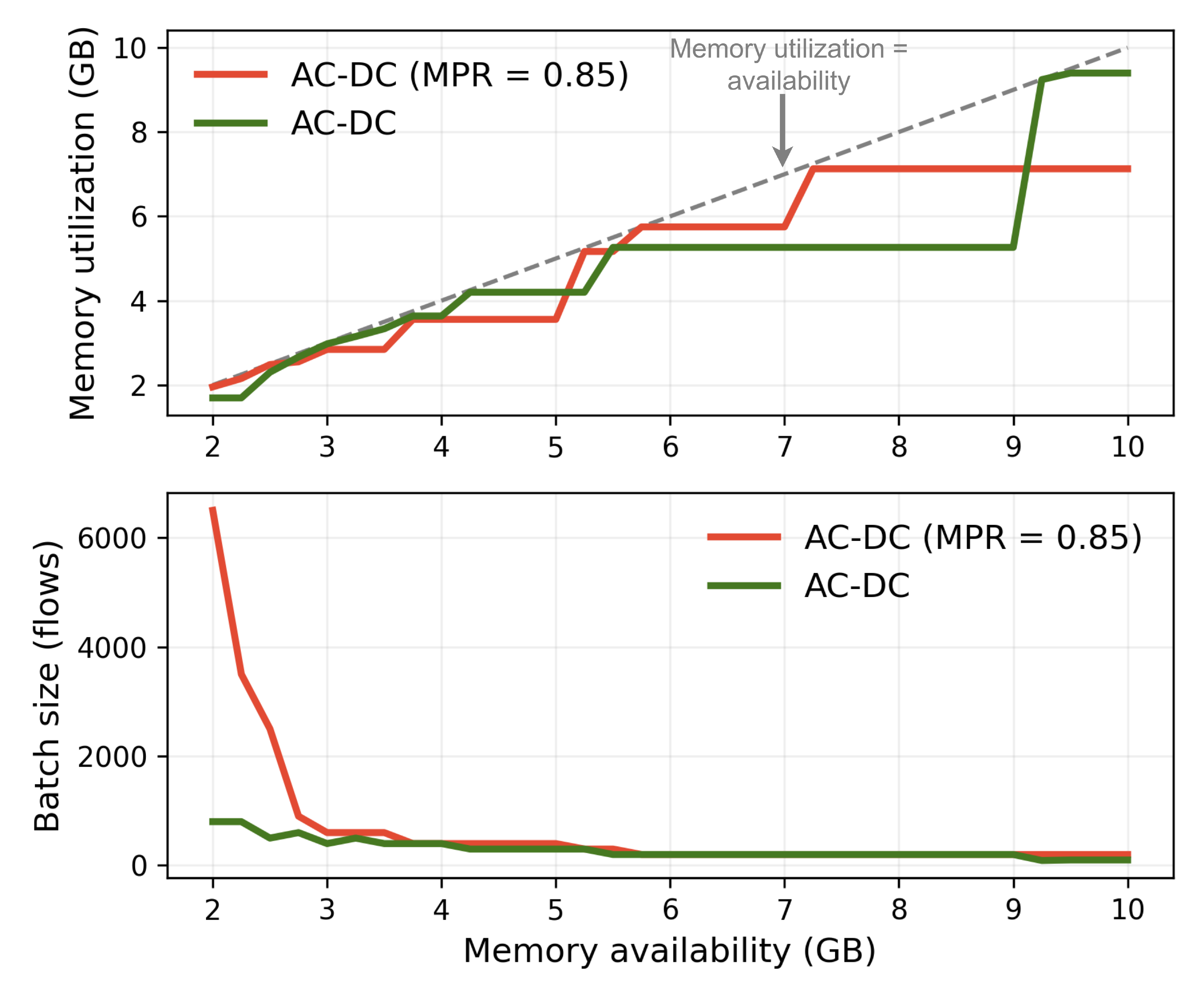}
	\caption{{\sysname} capitalizes on available memory by lowering
			the batch size; enforcing a minimum memory requirement
			increases the overall batch size selections.}
	\label{fig:dcm_fix_rate}
\end{figure}

As depicted in Figure~\ref{fig:dcm_fix_rate}, regardless of whether the minimum
performance requirement is set, \sysname{} exhibits gradual increases in memory
utilization and reductions in batch size as the system memory availability
increases. This behavior aligns with the adaptive  scheduler's objective to
maximize the performance-to-TTD ratio because reducing the batch size and
choosing classifiers with smaller feature requirements can both lead to a
decrease in TTD. However, to this behavior follows an increase in memory
requirements, as a reduced batch size results in a higher number of concurrent
classifiers.

We also observe that the selected batch sizes are generally larger when a
minimum performance requirement of 0.85 in F1-Score is introduced. In this case,
the classifier selection process becomes less flexible due to the necessity of
selecting classifiers with better performance and, consequentially, higher unit
memory requirement per classifier instance. As a result, the framework opts for
larger batch sizes to reduce the number of concurrently running instances and to
conform to the memory availability.

\begin{figure}[t]
	\centering     
 \includegraphics[width=0.85\columnwidth]{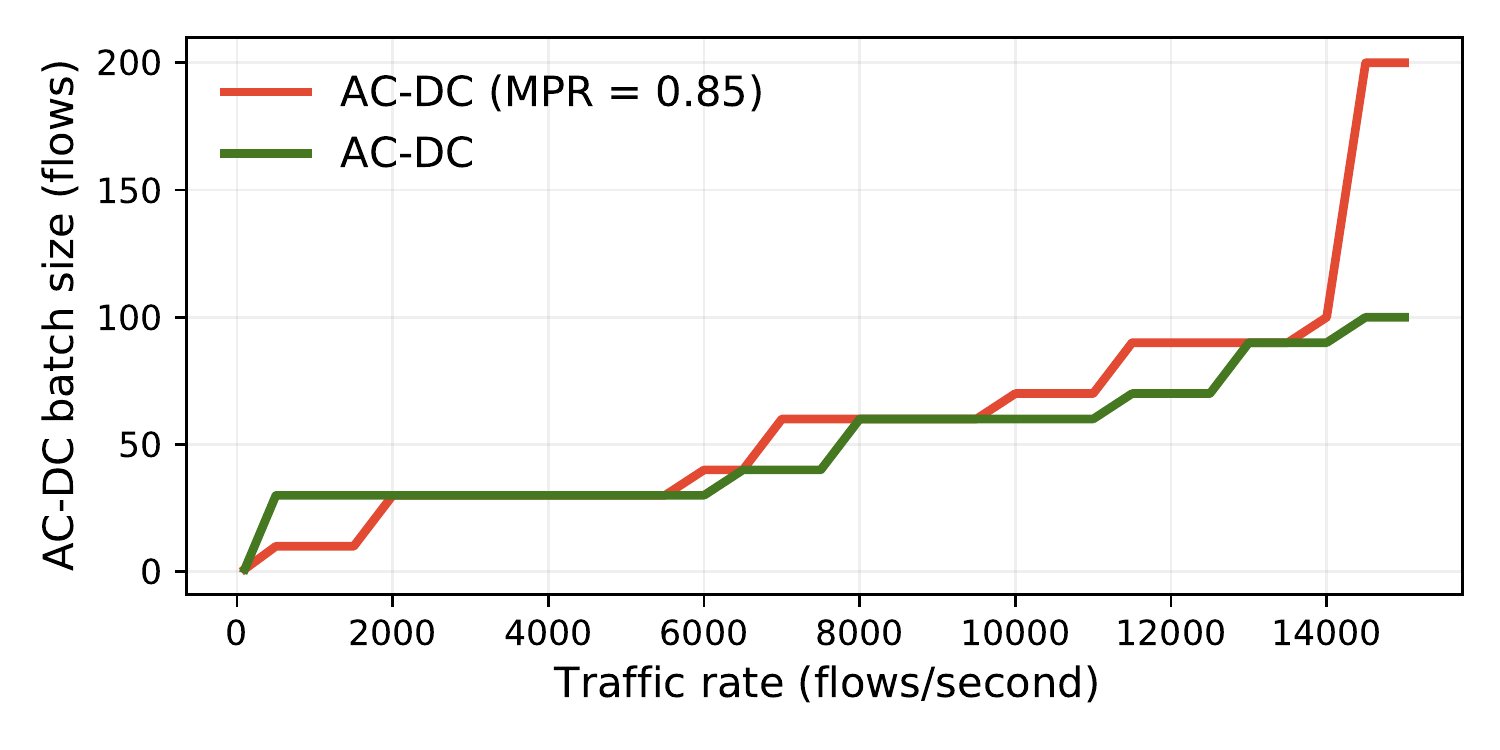}
	\caption{{\sysname} increases batch size facing increasing traffic rate
 to conform to insufficient memory availability.}
	\label{fig:dcm_fix_mem}
\end{figure}

\paragraph{{\sysname} Adjustments Facing Varying Traffic Rate.}
Following the same approach, we examine the effects of traffic rate on
{\sysname} by gradually increasing the traffic rate from 100 to
15,000 flows/second while fixing the memory availability at an indicative capacity of 10~GB (similar results can be observed at different memory capacities). As shown in Figure~\ref{fig:dcm_fix_mem}, the adaptive
classification framework increases its batch size as the traffic rate grows.
This confirms that the {\sysname} is operating as expected. Increasing traffic
rate continuously with no additional memory resource can result in an increased
number of concurrently running classifier instances that will eventually violate
the memory availability. As our adaptive scheduler imposes a hard constraint
on conforming to current memory availability, it addresses the issue by
increasing the batch size and thus reducing the number of concurrently running
classifier instances to meet the memory constraint. And, following the same line
of reasoning we previously analyzed, the batch sizes are comparatively higher
when a minimum performance requirement is added and this is to compensate for
the higher feature requirements and memory consumption in the selected
classifiers.


\section{Related Work}\label{sec:related}
Traffic classification has been a long-standing objective for researchers 
and practitioners. Early methods often relied on heuristics such as TCP/UDP 
port numbers and packet payload signatures~\cite{yoon2009internet, 
aceto2010portload, 4351725, sen2004accurate, moore2005toward}. With the 
availability of large network traffic datasets and increased computing 
power, machine learning-based traffic classification methods emerged, 
consisting of two prominent categories: flow-statistics-based and 
packet-capture-based.

\paragraph{Flow-Statistics-Based Methods.}
Methods based on flow statistics train models using important information from
network flows, such as the number and size of packets, flow duration, and
inter-arrival time~\cite{bernaille2006early, karagiannis2005blinc,
lang2003synthetic, lang2004synthetic}, or relying on flow statistics made
available by tools such as NetFlow~\cite{claise2004cisco} and
IPFIX~\cite{claise2008specification}. As a result, flow-statistics classifiers
have low TTD and low memory requirements. However, flow-statistics classifiers
are relatively brittle due to the constantly evolving nature of the
Internet~\cite{trevisan, moore2006inferring, shavitt2005dimes,
williamson2001internet, liu2021characterizing,
bronzino2021mapping,jiangtowards}. Changes over time can alter the landscape of
used traffic features, as highly engineered features extracted by domain experts
are static and hard to update, especially with increasing level of encryption
making them unavailable or ineffective. \sysname{} aims to achieve similar
efficiency of flow-statistics-based methods while overcoming these limitations.

\paragraph{Packet-Capture-Based Methods.}
With advancements in computing power and more sophisticated ML models, recent
studies have started applying representation learning to traffic
classification~\cite{bernaille2006early, bernaille2006traffic,
karagiannis2005blinc, lotfollahi2020deep, shapira2019flowpic, cui2019session,
sun2020encrypted, bu2020encrypted, zheng2022mtt}. This approach eliminates the
need for domain experts to identify and extract features. Packet-capture
classifiers are often more accurate and robust, even in the face of encryption,
as they can identify complex and predictive patterns in detailed packet-level
features. However, the higher classification performance comes at the cost of
increased efficiency overhead~\cite{cui2019session, sun2020encrypted,
bu2020encrypted, zheng2022mtt}. Raw network features must be preprocessed and
converted into formats compatible with ML models, which is time-consuming and
memory-intensive. \sysname{} aims to achieve similar performance to
packet-capture-based methods while overcoming these limitations.

\paragraph{Efficiency-Oriented Classifiers.}
There is a rich body of literature focused on developing low-latency network
traffic classifiers. Most of these classifiers aim to improve efficiency through
reducing model complexity, such as using lightweight models or simpler feature
representation techniques~\cite{koksal2022markov, crankshaw2017clipper,
li2020train, qiu2022traffic, tong2014high}. Other approaches utilize ensemble
techniques or tree-structured machine learning to improve classification
efficiency using diverse families of ML algorithms~\cite{devprasad2022context,
liu2019adaptive, farid2013adaptive}. These efforts primarily focus on the model
execution at the end of classification pipelines, with less emphasis on
manipulating and introducing flexibility into the feature space and
preprocessing phase. Another noticeable trend in the literature is the
relatively lower priority given to optimization and conformance to computational
resources, such as memory overheads, which can significantly impact the
classifier's practicality. And {\sysname} attempts addresses this issue by
capitalizing system memory availability without violating the system memory
constraint.
\section{Discussion and future work}\label{sec:future_work}
\paragraph{Beyond Packet Headers.}
In this study, we have limited our evaluation to packet header-level 
features for the purpose of demonstrating the feasibility of using 
adaptive ensemble classification in network scenarios. Although this 
approach has shown promising results, incorporating additional 
network features into the evaluation has the potential to further 
improve the classification accuracy and efficiency. These additional 
features may include finer-grained bit-level features, derived 
features, and cross-layer features. In future work, we plan to 
extend the feature space and test the robustness of our proposed 
framework to better understand its effectiveness, to meet the needs 
of various networked systems.
 
\paragraph{Ensemble Models.}
{\sysname} primarily focuses on addressing the bottleneck in 
preprocessing efficiency as compared to traditional ensemble-based 
approaches to aggregate multiple learning algorithms. However, it 
is also possible to enhance {\sysname} by incorporating different 
model types and architectures for different features. The flexible 
feature requirements of {\sysname} allow for the selection of 
machine learning models that are best suited to the chosen 
features, and incorporating a diverse set of model types and 
architerctures may lead to improved efficiency and performance. 

\paragraph{Encrypted Network Traffic.}
The widespread adoption of encryption techniques such as QUIC 
and VPN has created a challenge for ML 
classifiers, as they are often unable to interpret the content of 
the encrypted traffic. In light of this challenge, exploring the 
application of {\sysname} to network traffic under varying levels 
of encryption is a promising avenue of research. The adaptable 
feature requirements of {\sysname} may help to mitigate the risk
of overfitting to a limited set of features, which can be a 
concern when the traffic is increasingly become more encrypted.
\section{Conclusion}\label{sec:conclusion}
Classifying network traffic at varying traffic rates with limited resources is challenging 
for a spectrum of network management tasks. Classical solutions that use engineered flow 
statistics are quick but struggle with accuracy as the internet evolves, while 
recent representation learning is accurate but fails to be fast and memory efficient.
This paper introduces a new Adaptive Constraint-Driven Classification
framework that adjusts to system resource 
availability and incoming traffic rate, balancing performance and classification
throughput while conforming to memory availability.
The framework uses a heuristic-based feature
exploration algorithm to create a
pool of classifiers and choose the most appropriate classifier and batch size,
based on memory availability 
and traffic rate. The evaluation shows that the framework performs better (over 100\%
higher in F1-Score) than flow-statistics classifiers and has a much higher classification 
throughput (more than 150x) compared to conventional packet-capture
classifiers. We believe that the AC-DC framework can be employed beyond 
traffic identification and there's a rich research avenue in its integration to other 
networked systems on problems beyond classification.

\noindent \textbf{Ethics:} This work does not raise any ethical issues.
\bibliographystyle{ACM-Reference-Format}
\bibliography{sigcomm22}
\end{sloppypar}

\appendix
\newpage
\section*{Appendix}
\begin{figure}[thb]
	\centering     
	\includegraphics[width=0.85\columnwidth]{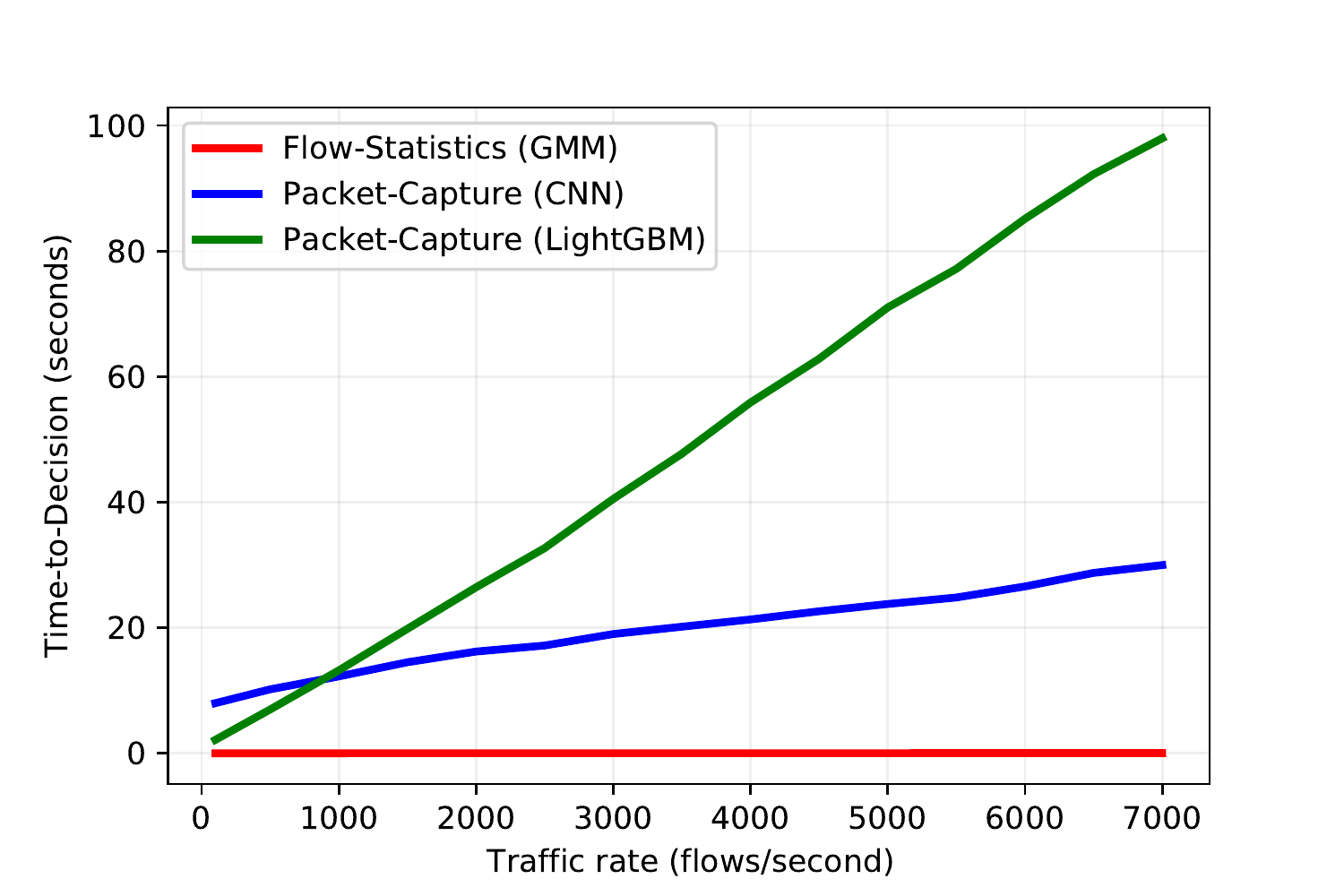}
	\caption{TTD comparison of the baseline classifiers with traffic rates
		from 10 to 7000 flows/second}
	\label{fig:ttd_compare_app}
\end{figure}
\begin{figure}[thb]
	\centering     
	\includegraphics[width=0.85\columnwidth]{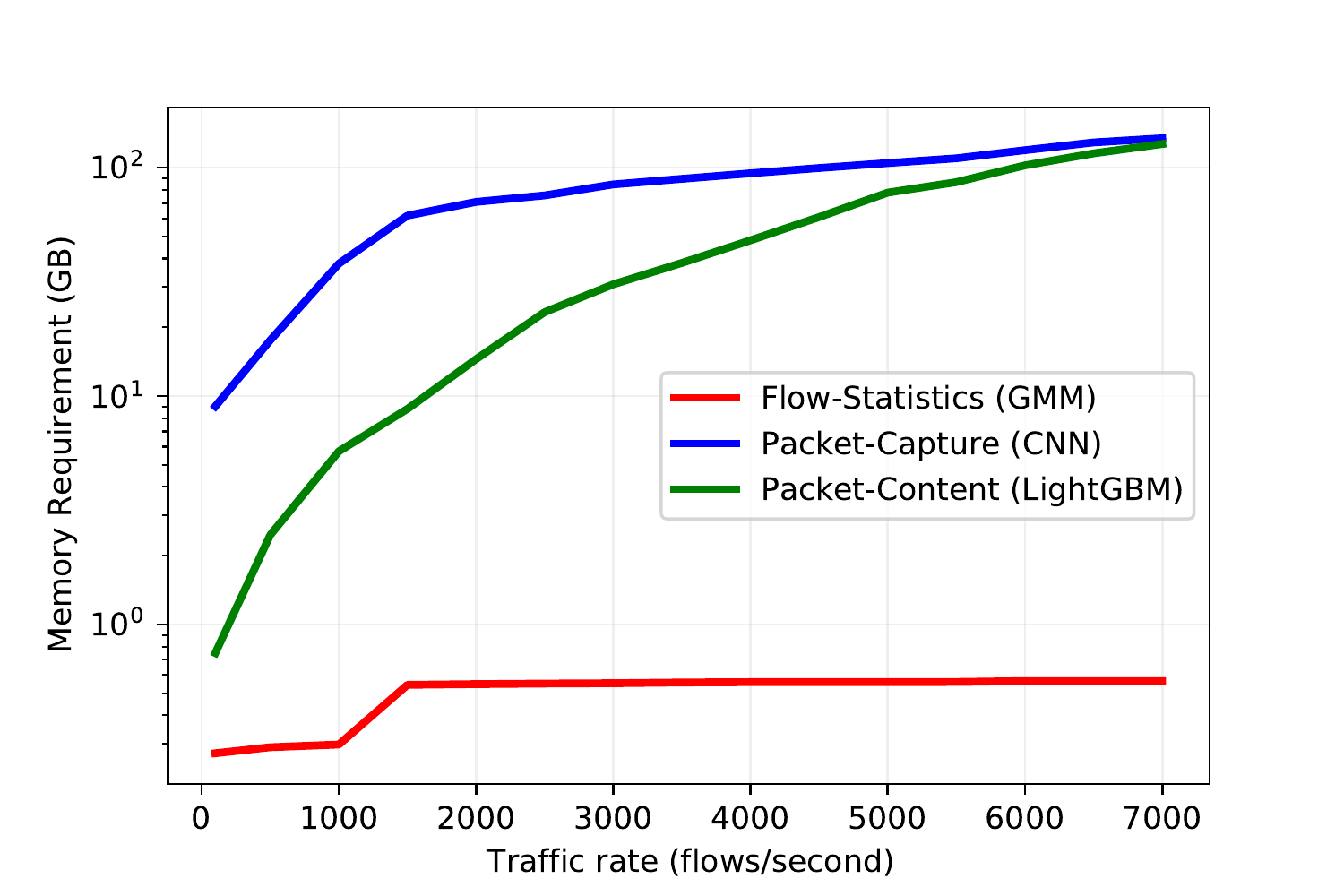}
\caption{Minimum memory requirement comparison of the baseline classifiers
	with traffic rates
	from 10 to 7000 flows/second}
	\label{fig:mem_compare_app}
\end{figure}
%
\begin{figure}[thb]
	\centering     
	\includegraphics[width=0.85\columnwidth]{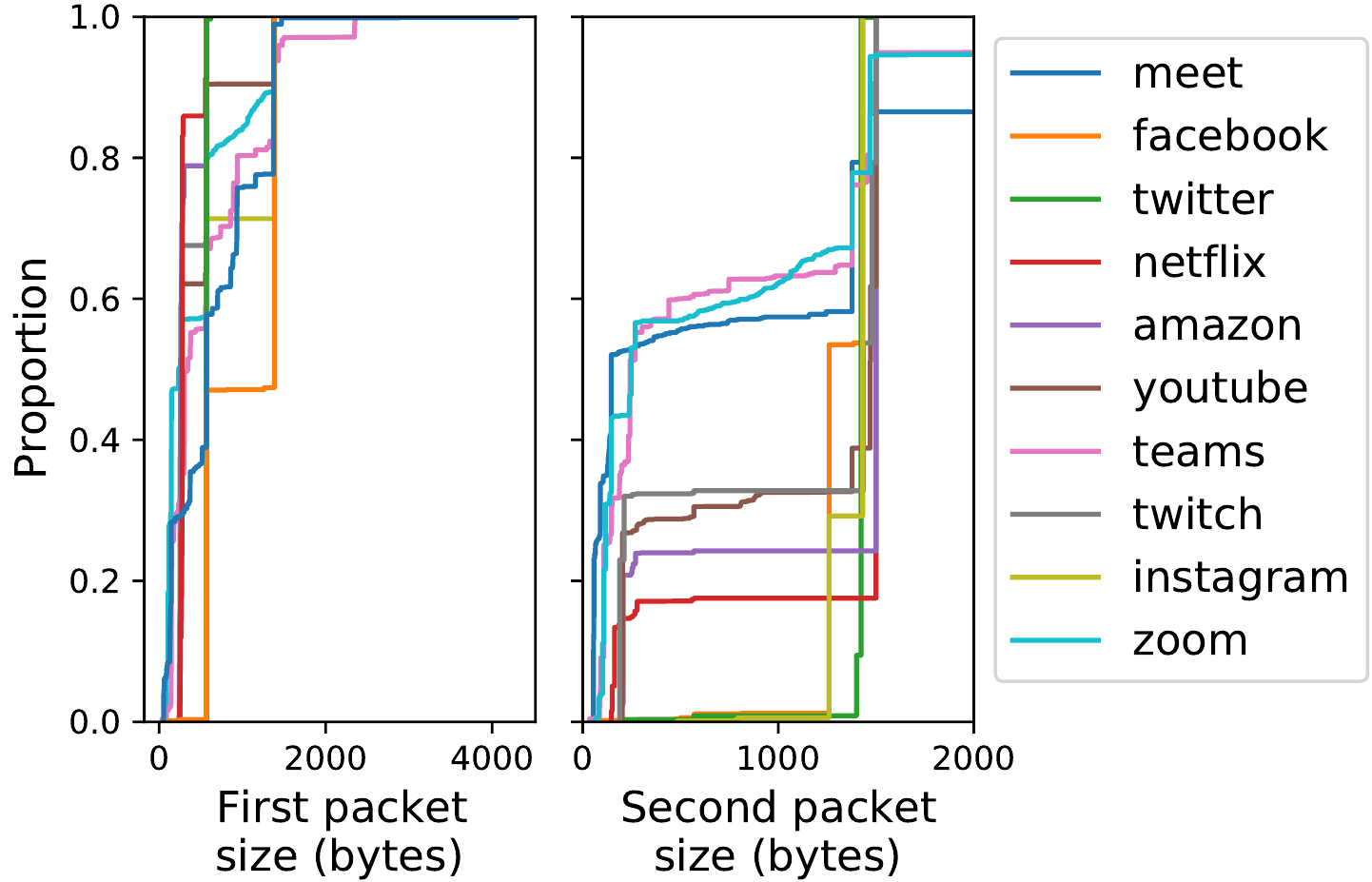}
	\caption{First and second packet size features used by the
		flow-statistics classifier show less distinctiveness across specific
		applications (\eg, Netflix vs Zoom)}
	\label{fig:gmmdistribution}
\end{figure}
%

\begin{table}[thb]
	\centering
	\resizebox*{\columnwidth}{!}{
		\begin{tabular}{|l|l|l|l|}
			\toprule
			\textbf{Header} & \textbf{Field} & \textbf{\makecell{Passing Preliminary\\Feature Selection}}  & \textbf{\makecell{Passing Heuristic\\Feature Selection}}\\\hline
			ipv4&ttl&Y&Y\\ \hline
			tcp&opt&Y&Y\\ \hline
			ipv4&dfbit&Y&Y\\ \hline
			tcp&doff&Y&Y\\ \hline
			tcp&wsize&Y&Y\\ \hline
			tcp&fin&Y&Y\\ \hline
			ipv4&cksum&Y&Y\\ \hline
			tcp&ackf&Y&Y\\ \hline
			udp&len&Y&Y\\ \hline
			tcp&cksum&Y&Y\\ \hline
			udp&cksum&Y&Y\\ \hline
			ipv4&tl&Y&Y\\ \hline
			ipv4&tos&Y&Y\\ \hline
			ipv4&proto&Y&Y\\ \hline
			tcp&seq&Y&Y\\ \hline
			tcp&psh&Y&Y\\ \hline
			tcp&ackn&Y&Y\\ \hline
			tcp&rst&Y&Y\\ \hline
			tcp&res&Y&N\\ \hline
			ipv4&foff&Y&N\\ \hline
			tcp&urp&Y&N\\ \hline
			tcp&urg&Y&N\\ \hline
			tcp&syn&Y&N\\ \hline
			tcp&ns&Y&N\\ \hline
			ipv4&hl&Y&N\\ \hline
			tcp&ece&Y&N\\ \hline
			ipv4&mfbit&Y&N\\ \hline
			ipv4&opt&Y&N\\ \hline
			ipv4&rbit&Y&N\\ \hline
			tcp&cwr&Y&N\\ \hline
			ipv4&ver&Y&N\\ \hline
			ipv4&id&Y&N\\ \hline
			ipv4&sport&N&N\\ \hline
			ipv4&dport&N&N\\ \hline
			ipv4&sip&N&N\\ \hline
			ipv4&dip&N&N\\ \hline
			tcp&payload&N&N\\ \hline
		\end{tabular}
	}
	\caption{Summary of the remaining features after preliminary selection.}
	\vspace{-5mm}
	\label{tab:feature_selection}
\end{table}

\begin{table}[thb]
	\centering
	\resizebox*{\columnwidth}{!}{
		\begin{tabular}{|l|l|l|l|}
			\toprule
			\textbf{Header} & \textbf{Field} & \textbf{\makecell{Number of Bits}}  & \textbf{\makecell{Feature Importance}}\\\hline
   ipv4&dfbit &    1   & 0.048111 \\ \hline
    tcp&fin    & 1   & 0.017778\\ \hline
   ipv4&ttl     &8   & 0.121000 \\ \hline
    tcp&doff     &4   & 0.032667\\ \hline
   tcp&ackf  &   1   & 0.008111 \\ \hline
   tcp&wsize &   16   & 0.028556 \\ \hline
      tcp&psh &    1  &  0.001333\\ \hline
  ipv4&cksum   & 16   & 0.009889\\ \hline
    udp&len   & 16   & 0.007778 \\ \hline
  tcp&cksum   & 16   & 0.007000\\ \hline
     ipv4&tl   &  8  &  0.003444 \\ \hline
    tcp&opt  & 320   & 0.107000 \\ \hline
   udp&cksum  &  16  &  0.005222 \\ \hline
   ipv4&tos    & 8   & 0.002333 \\ \hline
   ipv4&proto  &   8  &  0.002222\\ \hline
     tcp&rst    & 1   & 0.000111 \\ \hline
     tcp&seq   & 32   & 0.001444\\ \hline
    tcp&ackn   & 32   & 0.000333\\ \hline
		\end{tabular}
	}
	\caption{Summary of the remaining features after heuristics-based feature selection for
		pool of classifier generation.}
	\vspace{-5mm}
	\label{tab:final_feature_selection}
\end{table}

%

\begin{table}[thb]
	\centering
	\resizebox*{\columnwidth}{!}{
		\begin{tabular}{|l|l|l|l|}
			\toprule
			\textbf{Feature Requirements} & \textbf{\makecell{TTD\\(seconds)}} & \textbf{\makecell{Memory \\Requirement\\(MB per instance)}}  & \textbf{\makecell{Performance\\(F1-Score)}}\\\hline
ipv4-dfbit\&tcp-fin\&ipv4-ttl\&tcp-ackf&0.303&315&0.744\\\hline
ipv4-dfbit\&tcp-fin\&ipv4-ttl\&tcp-doff\&tcp-ackf&0.318&323&0.772\\\hline
ipv4-dfbit\&tcp-fin\&ipv4-ttl\&tcp-doff&0.32&317&0.773\\\hline
ipv4-dfbit\&tcp-fin\&ipv4-ttl\&tcp-doff\&tcp-psh&0.326&321&0.774\\\hline
ipv4-dfbit\&tcp-fin\&ipv4-ttl\&tcp-doff\&tcp-ackf\&tcp-psh&0.334&323&0.791\\\hline
ipv4-dfbit\&tcp-fin\&ipv4-ttl&0.293&317&0.693\\\hline
ipv4-dfbit\&tcp-fin\&ipv4-ttl\&tcp-psh&0.307&321&0.716\\\hline
ipv4-dfbit\&tcp-fin\&tcp-wsize&0.316&321&0.736\\\hline
ipv4-dfbit\&ipv4-ttl&0.282&331&0.655\\\hline
ipv4-dfbit\&tcp-fin\&ipv4-ttl\&tcp-wsize&0.357&317&0.826\\\hline
ipv4-dfbit\&tcp-fin\&ipv4-ttl\&tcp-doff\&ipv4-tos&0.352&323&0.797\\\hline
ipv4-dfbit\&tcp-fin\&ipv4-ttl\&tcp-doff\&tcp-ackf\&ipv4-tos&0.354&317&0.792\\\hline
ipv4-dfbit\&tcp-fin\&ipv4-ttl\&ipv4-tl&0.362&319&0.8\\\hline
ipv4-dfbit\&tcp-fin\&ipv4-ttl\&tcp-doff\&udp-cksum&0.373&338&0.815\\\hline
ipv4-dfbit\&tcp-fin\&ipv4-ttl\&tcp-doff\&tcp-ackf\&udp-cksum&0.376&312&0.815\\\hline
		\end{tabular}
	}
	\caption{Top 15 classifiers in the pool of classifiers
		generated for evaluation purpose, with the highest
		performance-to-efficiency ratios
		at a selected batch size
		of 500 flows.}
	\vspace{-5mm}
	\label{tab:evaluation_classifiers}
\end{table}

%

\begin{figure}[thb]
	\centering     
	\includegraphics[width=\columnwidth]{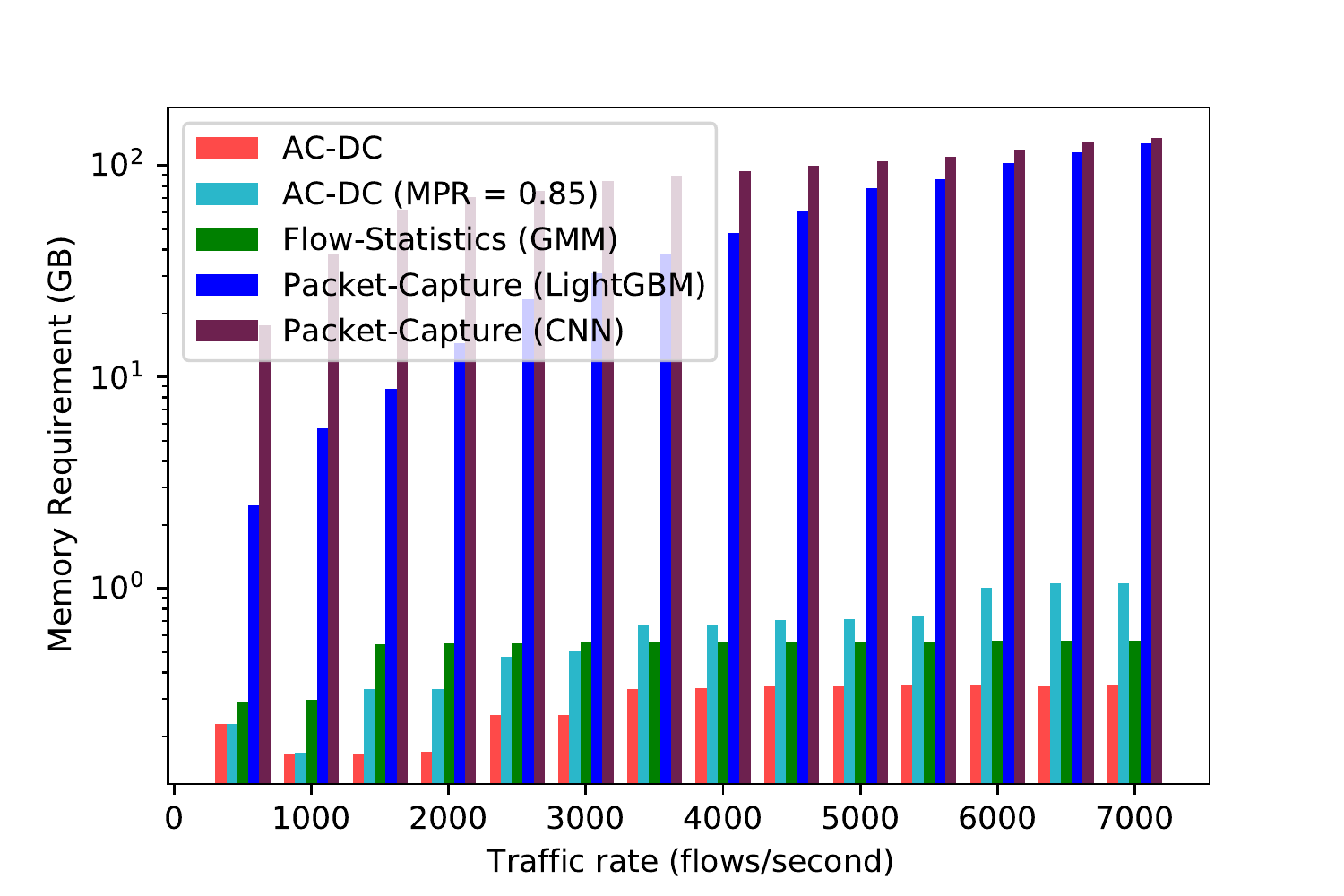}
	\caption{Minimum memory requirement comparison of AC-DC against conventional classifiers
		with traffic rates
		from 500 to 7000 flows/second}
	\label{fig:eval_mem_compare_app}
\end{figure}

\end{document}